\documentclass[aps,floats,twocolumn]{revtex4}
\usepackage{tabularx,graphicx}
\usepackage{epsfig}
\usepackage{amsmath}
\usepackage{bbm}
\usepackage{graphicx}

\begin{document}
%\draft
%\flushbottom
%\twocolumn[
%\hsize\textwidth\columnwidth\hsize\csname @twocolumnfalse\endcsname

\title{Renormalization group approach to chiral symmetry breaking in graphene}
\author{J. Gonz\'{a}lez \\}
\address{Instituto de Estructura de la Materia,
        Consejo Superior de Investigaciones Cient\'{\i}ficas, Serrano 123,
        28006 Madrid, Spain}

\date{\today}

\begin{abstract}
%\widetext
We investigate the development of a gapped phase in the field theory of Dirac 
fermions in graphene with long-range Coulomb interaction. In the large-$N$ 
approximation, we show that the chiral symmetry is only broken below a critical 
number of two-component Dirac fermions $N_c = 32/\pi^2$,
that is exactly half the value found in quantum electrodynamics in 2+1
dimensions. Adopting otherwise a ladder approximation, we give evidence of the 
existence of a critical coupling at which the anomalous dimension of the order
parameter of the transition diverges. This result is consistent with the 
observation that chiral symmetry breaking may be driven by the long-range 
Coulomb interaction in the Dirac field theory, despite the divergent scaling 
of the Fermi velocity in the low-energy limit.

\end{abstract}
%\pacs{71.10.Pm,73.22.-f}

%]
\maketitle

%\narrowtext
%\tightenlines

%\newpage

\section{Introduction}

The fabrication of single layers of carbon with atomic thickness has provided
us with a laboratory to explore new physics\cite{geim,kim}, as the electrons 
in this so-called 
graphene behave at low energies as massless Dirac fermions, displaying 
conical valence and conduction bands\cite{rmp}. Apart from its quite 
interesting properties from the applied point of view, the new material 
offers the possibility of studying an electron system that is a variant of 
quantum electrodynamics (QED) in the strong coupling regime, with unusual
features as shown for instance in Refs. \onlinecite{nil,fog,shy,per,ter}. 
 
A remarkable feature of this field theory of electrons in graphene is 
its scale-invariant character\cite{np2}. This means for practical purposes 
that, while many-body corrections give rise in general to dependences on the 
high-energy cutoff, these are susceptible of being reabsorbed into
the definition of physical quantities. Consequently, some of the parameters
of the theory may have a nontrivial scaling in the low-energy
limit. The quasiparticle weight is for instance renormalized, and it would be 
driven to zero if its flow were not arrested by the divergence of the Fermi 
velocity in the infrared\cite{prbr}. This marginal behavior leaves anyhow an 
imprint in the quasiparticle decay rate\cite{unconv}, with an unconventional
dependence on energy which has been observed experimentally\cite{xu}. 
 
An important phenomenon that may take place in a system of massless Dirac 
fermions is the opening of a gap in the regime of strong interaction. In this 
respect, the case of QED in 2+1 dimensions can serve as a good example, in 
which the original $U(N)$ chiral symmetry of the theory with $N$ massless 
two-component Dirac fermions is spontaneously broken below a critical number 
of flavors $N_c$ \cite{appel}. This chiral symmetry breaking (CSB) has been 
also studied in graphene by a number of 
analytical\cite{khves,gus,ale,son,her,jur} as well as numerical 
methods\cite{drut1,drut2,hands,arm}. The conclusion to be drawn 
from different approaches is that a gap can open up in the Dirac spectrum, 
though the effect may only appear below some critical value of $N$ and above 
some critical interaction strength. In this picture, there remain however 
important questions to be addressed, related to the effect of the above 
mentioned scaling of the parameters in the model. We point out in particular 
that the strength of any four-fermion interaction in the Dirac field theory 
has to be measured relative to the weight of the kinetic energy, that scales 
with the Fermi velocity. Then, it is crucial to clarify whether the divergence 
of this parameter in the infrared may prevent the CSB even for a small number
of Dirac fermions.

In this paper we apply renormalization group methods to study 
the CSB in the field theory of Dirac fermions in graphene. 
We consider that this electron system is governed at low energies
by the hamiltonian
\begin{eqnarray}
\lefteqn{H = i v_F \int d^2 r \; \overline{\psi}_i({\bf r}) 
 \mbox{\boldmath $\gamma   \cdot \nabla $} \psi_i ({\bf r}) }   \nonumber \\
  &   &     + \frac{e^2}{8 \pi} \int d^2 r_1
\int d^2 r_2 \; \rho ({\bf r}_1) 
       \frac{1}{|{\bf r}_1 - {\bf r}_2|} \rho ({\bf r}_2)  \;\;\;\;\;
\label{ham}
\end{eqnarray}
where  $\{ \psi_i \}$ is a collection of $N/2$ four-component Dirac 
spinors, $\overline{\psi}_i = \psi_i^{\dagger} \gamma_0 $, and 
$\rho ({\bf r}) = \overline{\psi}_i ({\bf r}) \gamma_0 \psi_i ({\bf r})$.
The matrices $\gamma_{\sigma } $ satisfy 
$\{ \gamma_\mu, \gamma_\nu \} = 2 \: {\rm diag } (1,-1,-1)$
and can be conveniently represented in terms of Pauli matrices as
$\gamma_{0,1,2} = (\sigma_3, \sigma_3 \sigma_1, \sigma_3 \sigma_2) \otimes
 \sigma_3$, where the first factor acts on the two sublattice components of 
the graphene lattice. 
%Such effective field theory provides a good starting point for a RG
%analysis since, given that the scaling dimension of the
%$\Psi({\bf r})$ field is $-1$ (in length units), the four-fermion 
%Coulomb interaction turns out to be scale invariant, at this level, with a 
%dimensionless coupling constant $e^2$.
Our aim is to elucidate whether a term of the type 
\begin{equation}
\rho_m ({\bf r}) =  \overline{\psi}({\bf r}) \psi ({\bf r}) 
\end{equation} 
is generated spontaneously in the hamiltonian 
of the electron system. A convenient way to address this question is to look 
at the susceptibility built from that operator, that is, at the correlator
\begin{equation}
\Pi ({\bf q},\omega ) = i \int_{-\infty }^{+\infty } dt \;  e^{i \omega t}
  \;  \langle T \rho_m ({\bf q}, t) \rho_m (-{\bf q}, 0) \rangle
\label{ms}
\end{equation}
$\Pi ({\bf q},\omega )$ is actually a response function measuring the
reaction of the system under a slight difference of scalar potential 
in the two sublattices of the graphene lattice. 
A divergence of $\Pi ({\bf 0}, 0 )$ at some particular value of the coupling 
constant can be interpreted as the signal that $\rho_m $ is getting 
a nonvanishing expectation value, which is in turn the signature of the 
opening of a gap in the Dirac spectrum.

We will take advantage of the power of the renormalization group to 
characterize the possible singular behavior of $\Pi $ as a function of 
$e^2/v_F$. For this purpose, we concentrate on the corrections to the vertex
built from $\overline{\psi} \psi $, as shown in Figs. \ref{one} and \ref{three}. 
%In the case of an operator like $\rho_m $ which is not present in the 
%hamiltonian, the cutoff dependence of its correlators is to be reabsorbed 
%into the scale $Z_{\psi^2}$ of the composite operator. This implies that 
In the process of renormalization, 
$\rho_m  $ may get in general an anomalous dimension $\gamma_{\psi^2}$, 
modifying the naive scaling of the susceptibility,
\begin{equation}
\Pi ({\bf q}, 0) \sim  |{\bf q}|^{1 - 2\gamma_{\psi^2}}
\label{scal}
\end{equation}
%The relation between the scale of the composite operator and the anomalous
%dimension is given by 
%\begin{equation}
%\gamma_{\psi^2} = -\Lambda 
% \frac{\partial \log Z_{\psi^2} }{\partial \Lambda }
%\label{anom}
%\end{equation}
%where $\Lambda $ is the high-energy cutoff of the theory. We will focus then
In what follows, we apply different approaches
for the determination of $\gamma_{\psi^2}$, in order to establish
the existence of a singular behavior in the long-distance scaling of the 
susceptibility $\Pi $.

\section{Large-N approximation}

\begin{figure}
\begin{center}
\mbox{\epsfxsize 3.0cm \epsfbox{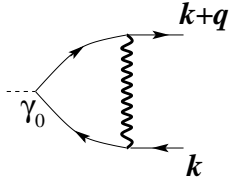}}
\end{center}
\caption{Quantum corrections to the vertex built from 
$\overline{\psi} \psi $ in the large-$N$ 
approximation, where the interaction between electrons is taken as the RPA 
dressed Coulomb potential (thick wavy line).}
\label{one}
\end{figure}

We can go beyond the usual perturbative approach in the coupling $e^2/v_F$ by 
taking formally a large number $N$ of fermion flavors, to perform then the sum 
of all the diagrams that arise to leading order in a $1/N$ expansion.
If we think of all possible contributions to the expectation value 
$\langle \rho_m ({\bf q}) \psi ({\bf k}+{\bf q}) \psi^{\dagger} ({\bf k}) 
\rangle$, it is clear that the leading corrections in $1/N$ are given
by the iteration of the exchange of 
electron-hole bubbles in the interaction between the $\psi $ and 
$\psi^{\dagger}$ fields. This amounts to adopt the RPA for the 
dressed Coulomb interaction represented in Fig. \ref{one}. Introducing
the polarization $\chi ({\bf q}, \omega_q)$, we get for the corresponding
vertex function
\begin{eqnarray}
\lefteqn{   \Gamma ({\bf q}; {\bf k}) =  \gamma_0 +  i \sum_{n=0}^{\infty }  
  \int \frac{d^2 p}{(2 \pi )^2} \frac{d \omega_p}{2 \pi }
  G_0({\bf p}, \omega_p) \gamma_0       }                      \nonumber       \\
 &  &   \;\;\;\;\;   
             G_0({\bf p}+{\bf q}, \omega_p)   \frac{e^2}{2|{\bf p}-{\bf k}|} 
  \left( \frac{e^2 \chi ({\bf p}-{\bf k}, \omega_p-\omega_k)}
                {2|{\bf p}-{\bf k}|} \right)^n    \;\;\;\;
\label{vert}
\end{eqnarray}
where $G_0$ stands for the free Dirac propagator. We recall that, in the case
of $N$ two-component Dirac fermions, $\chi ({\bf q}, \omega_q) = -(N/16) 
 {\bf q}^2/\sqrt{v_F^2 {\bf q}^2 - \omega_q^2}$.

An interesting feature of the sum in (\ref{vert}) is that, while a 
high-energy cutoff $\Lambda $ has to be imposed to make the integrals finite, 
all the terms show the same degree of logarithmic dependence on the cutoff. 
By passing to imaginary frequency $i \overline{\omega}_p = \omega_p$, we can 
compute the divergent contribution to the vertex as
\begin{eqnarray}
\Gamma  
&  \approx  &  \gamma_0  +  \gamma_0 \frac{e^2}{2}  \sum_{n=0}^{\infty }  
       (-1)^n  \left(\frac{N e^2}{32} \right)^n         \nonumber       \\
 & & \times \int \frac{d^2 p}{(2 \pi )^2} \frac{d \overline{\omega}_p}{2 \pi }      
  \frac{|{\bf p}|^{n-1} }{(\overline{\omega}_p^2 + v_F^2 {\bf p}^2)^{n/2+1} }
                                                          \nonumber     \\
&  \approx  &   \gamma_0  +  \gamma_0  \frac{4}{\pi^{3/2}N}   
   \sum_{n=0}^{\infty } (-1)^n g^{n+1} 
   \frac{\Gamma(\frac{1}{2}+\frac{n}{2})}{\Gamma(1+\frac{n}{2})}  
 \int^{\Lambda }  \frac{d (v_F |{\bf p}| )}{v_F |{\bf p}| } \nonumber   \\
  &  \approx  &  \gamma_0  +   \gamma_0  \frac{8}{\pi^2 N} g 
   \frac{\arccos (g)}{\sqrt{1-g^2}}  \log \Lambda
\label{div}
\end{eqnarray}
where $g = (N/32)e^2/v_F$. We note that the singularity at $g = 1$ is only 
apparent, as the function in (\ref{div}) can be continued analytically 
to $g > 1$ by taking 
$\arccos (g) = i \log(g + \sqrt{g^2-1})$. In this way, we end up 
with an expression of the vertex that becomes sensible for arbitrarily large 
values of the effective coupling.

The divergence of the vertex $\Gamma $ at large $\Lambda $ has to be removed 
by absorbing the dependence on the cutoff into the scale $Z_{\psi^2}$ of the 
composite field $\overline{\psi} \psi $\cite{amit}. However, this is not the 
only field redefinition to be accomplished, as the finiteness of the full 
Dirac propagator demands the introduction of a cutoff-dependent scale for the 
Dirac field, such that $\psi (\Lambda ) = Z_{\psi}^{1/2} (\Lambda ) \psi_{\rm ren}$.
The electron self-energy can be actually found in Ref. \onlinecite{prbr} 
to dominant order in the $1/N$ approximation, providing the result
%\begin{eqnarray}
%\lefteqn{  \frac{1}{G}   =   \frac{1}{G_0} - \Sigma        }        \nonumber    \\
%   &  &   \approx   \;  Z_{\psi} 
%   (\omega_k - Z_v v_F  \mbox{\boldmath $\sigma \cdot$} {\bf k})     \nonumber  \\
%   &  &  - Z_{\psi}  \omega_k  \frac{8}{\pi^2 N}  
%  \left( 2 +  \frac{2 -  g^2}{g} 
%  \frac{\arccos g}{ \sqrt{1-g^2}}  - \frac{\pi}{g}  \right)   \log \Lambda   
%                                                                  \nonumber \\
%   &  &    + Z_{\psi} v_F  \mbox{\boldmath $\sigma \cdot$} {\bf k} 
% \frac{8}{\pi^2 N} 
% \left( 1 + \frac{\sqrt{1-g^2}}{g} \arccos g - \frac{\pi}{2g} \right) 
%     \log \Lambda   \;\;\;\;\;\;
%\label{selfe}
%\end{eqnarray}
%From this expression, we find that $Z_{\psi}$ has to be adjusted to 
\begin{equation}
Z_{\psi}   \approx   1 + \frac{8}{\pi^2 N}    \left( 2 +  \frac{2 -  g^2}{g} 
   \frac{\arccos g}{ \sqrt{1-g^2}} - \frac{\pi}{g} \right)  \log \Lambda   
\end{equation}

The cutoff-independence of the vertex $\Gamma $ must be guaranteed
after multiplication by $Z_{\psi}$ and the scale $Z_{\psi^2}$ of the composite 
field. We define the renormalized vertex as 
$\Gamma_{\rm ren} = Z_{\psi^2} Z_{\psi} \Gamma $. By imposing the finiteness
of $\Gamma_{\rm ren}$, we obtain to leading order in the $1/N$ expansion
\begin{equation}
Z_{\psi^2}   \approx   1 - \frac{8}{\pi^2 N}    \left( 2 +  \frac{2}{g} 
   \frac{\arccos g}{ \sqrt{1-g^2}} - \frac{\pi}{g} \right)  \log \Lambda  
\label{largen}
\end{equation}

The knowledge of $Z_{\psi^2}$ can now be used to determine the anomalous 
scaling of the susceptibility $\Pi ({\bf q}, \omega)$. This correlator involves 
two composite operators $\overline{\psi} \psi $, and it can be made 
cutoff-independent by multiplying each of them by their 
renormalization factor. The finite susceptibility is then
$\Pi_{\rm ren} ({\bf q}, \omega) = Z_{\psi^2}^2 \Pi ({\bf q}, \omega) $.
A renormalization group equation can be obtained for $\Pi $, relying on the 
independence of the susceptibility on $\Lambda $ in the renormalized 
theory\cite{amit}. We obtain from the invariance of $\Pi_{\rm ren}$
\begin{equation}
\left( \Lambda \frac{\partial }{\partial \Lambda } 
  + \beta (g) \frac{\partial }{\partial g}   - 2 \gamma_{\psi^2} \right)
\Pi ({\bf q}, \omega) = 0  
\label{rge}
\end{equation}
with the anomalous dimension
\begin{equation}
\gamma_{\psi^2} = -\Lambda 
 \frac{\partial \log Z_{\psi^2} }{\partial \Lambda }
\label{anom}
\end{equation}
and $\beta (g) = \partial g/\partial \log \Lambda $.
In the bare theory with a cutoff $\Lambda $, it follows from dimensional 
analysis that the susceptibility $\Pi $ can be written in terms of a 
dimensionless function $\Phi (x)$ as 
$\Pi ({\bf q},0) = (|{\bf q}|/v_F) \Phi (v_F |{\bf q}| /\Lambda )$. 
Then, neglecting in a first approximation the scaling of 
the effective coupling, the solution of (\ref{rge}) implies that
$\Pi ({\bf q}, 0) \sim (|{\bf q}|/v_F) 
   ( v_F |{\bf q}|/\Lambda )^{-2 \gamma_{\psi^2}}$,
with the behavior anticipated in Eq. (\ref{scal}).

\begin{figure}[t]
\begin{center}
\mbox{\epsfxsize 5.0cm \epsfbox{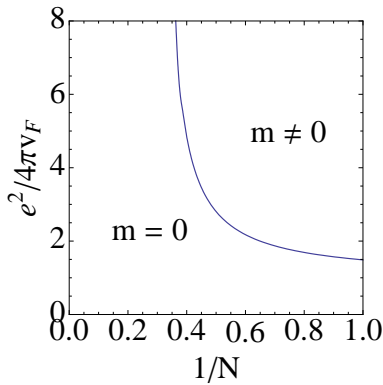}}
\end{center}
\caption{Phase diagram obtained to leading order in the $1/N$ approximation, 
showing the regime with massless Dirac fermions ($m = 0$) and the phase
with CSB ($m \neq 0$).}
\label{two}
\end{figure}

The anomalous dimension obtained from (\ref{largen}) is 
\begin{equation}
\gamma_{\psi^2} = \frac{8}{\pi^2 N}    \left( 2 +  \frac{2}{g} 
   \frac{\arccos g}{ \sqrt{1-g^2}} - \frac{\pi}{g} \right)
\label{gamma2}
\end{equation}
and it turns out to be a monotonous, increasing function of $g$. This means 
that, provided that it gets sufficiently large, there may exist a critical 
value $g_c$ at which $\Pi ({\bf q}, 0)$ becomes singular in the limit
${\bf q} \rightarrow 0$. The divergence of this susceptibility implies a 
long-wavelength instability, which can be interpreted as the development of
a nonvanishing expectation value of $ \overline{\psi} \psi $. On the 
other hand, the value of $g_c$ depends in general on the number of 
flavors $N$. We can draw then a boundary marking the onset of CSB 
in $(N,g)$ space. This line of transition, characterized by the 
condition $1-2\gamma_{\psi^2} = 0$, is shown in Fig. \ref{two}.

The expression (\ref{gamma2}) leads to the existence of 
a critical number of flavors $N_c$, above which CSB cannot 
take place. If we let $g \rightarrow \infty $ in that equation, we approach 
the maximum value of $\gamma_{\psi^2}$, from which we find $N_c = 32/\pi^2$.
It is very suggestive that this critical $N$ is precisely half the value 
obtained in QED in 2+1 dimensions\cite{appel}. 
Technically, the methods used to derive 
$N_c$ in each model cannot be easily compared, as QED is not a scale-invariant 
field theory in that number of dimensions. On intuitive grounds, however, 
one can understand the relation between the two values of $N_c$, as the
photon propagating in QED has two different degrees of freedom. This may 
explain that twice the number of flavors are needed there to equally screen 
the interaction, in comparison to our model with just 
the scalar Coulomb potential.

\section{Ladder approximation}

\begin{figure}
\begin{center}
\mbox{\epsfxsize 8.0cm \epsfbox{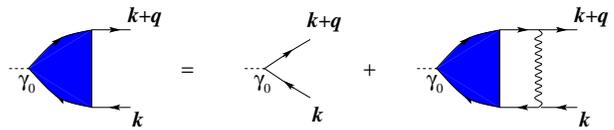}}
\end{center}
\caption{Self-consistent diagrammatic equation for the vertex 
$\Gamma ({\bf q}; {\bf k})$, equivalent to the sum of ladder diagrams built from
the iteration of the bare Coulomb interaction (thin wavy line).}
\label{three}
\end{figure}

We resort now to an approach that can better 
capture the behavior of the system to the right of the phase diagram of Fig. 
\ref{two}. For this purpose, it is pertinent to adopt a self-consistent 
approximation in the calculation of the vertex $\Gamma $, equivalent to the 
sum of ladder diagrams, by which the most divergent 
logarithmic dependences are taken into account at each perturbative 
level\cite{mis}. The approach is encoded in the self-consistent equation shown 
in Fig. \ref{three}. The perturbative solution leads to a power series in the 
effective coupling $\lambda \equiv e^2/8 \pi v_F $, where the term 
of order $\lambda^n$ diverges in general with the high-energy cutoff as 
$\log^n (\Lambda )$. The important point is that the 
set of diagrams considered in this way allows to implement a consistent 
renormalization of the theory, where $Z_{\psi^2}$ is free of 
nonlocal divergences, making possible a precise computation of the anomalous 
dimension $\gamma_{\psi^2}$.

A solution of the equation in Fig. \ref{three} has been given in
Ref. \onlinecite{fer} regularizing the momentum integrals with an infrared and
a high-energy cutoff. Here, 
in order to facilitate the calculation of the divergences of the vertex 
$\Gamma $, we define instead the field theory by analytic 
continuation to spatial dimension $d = 2 - \epsilon $. After integration in
the frequency variable, the self-consistent 
equation for the vertex takes the form
\begin{equation}
\Gamma ({\bf 0}; {\bf k}) = \gamma_0 + 2\pi \lambda_0 
    \int \frac{d^d p}{(2\pi )^d} \Gamma ({\bf 0}; {\bf p}) 
            \frac{1}{|{\bf p}|} \frac{1}{|{\bf p}-{\bf k}|}
\label{self}
\end{equation}
where the dimensionful coupling $\lambda_0$ is given in terms of an 
auxiliary momentum scale $\rho$ 
by $\lambda_0 = \lambda \rho^{\epsilon} $ (see below). Thus, powers of 
$\log \Lambda $ are traded by poles at $\epsilon = 0$ in the different 
perturbative contributions, which are easier to compute. In principle, Eq. 
(\ref{self}) could also afford a nonperturbative resolution, but the 
computation of the anomalous dimension would be complicated then as this 
is obtained from the residue of the $1/\epsilon $ pole. Unfortunately, a 
closed equation for that quantity cannot be written from Eq. (\ref{self}), 
which couples the equations for the coefficients of the different powers of
$\epsilon $. This is otherwise a 
natural consequence of the regularization of the diagrams, since 
the interdependence of the different poles is a key property of a 
renormalizable theory, as we illustrate below.

We resort then to an iterative resolution of Eq. (\ref{self}), by which we
can obtain a recursion between consecutive orders in the power series for 
the vertex
\begin{equation}
\Gamma ({\bf 0}; {\bf k}) = 
    \gamma_0 \left(1 + \sum_{n=1}^{\infty} \lambda_0^n \Gamma_n ({\bf k}) \right)
\end{equation}
It can be easily seen that the momentum dependence of the different orders takes in
general the form 
\begin{equation}
\Gamma_n ({\bf k}) = \frac{a_n }{|{\bf k}|^{n\epsilon}}
\end{equation}
Inserting
the $n$-th term of the series in the right-hand-side of Eq. (\ref{self}), we get
\begin{eqnarray}
  \frac{a_{n+1} }{|{\bf k}|^{(n+1)\epsilon}  }  & = &
 2\pi \int \frac{d^d p}{(2\pi )^d} \frac{a_n }{|{\bf p}|^{n\epsilon} }
            \frac{1}{|{\bf p}|} \frac{1}{|{\bf p}-{\bf k}|}     \nonumber   \\
  & = &  2\pi a_n \int \frac{d^d p}{(2\pi )^d} 
  \frac{\Gamma (1+\frac{n\epsilon}{2})}{\sqrt{\pi } \Gamma (\frac{1+n\epsilon}{2}) }
                                                               \nonumber     \\ 
  &   &  \times  \int_0^1  dx  \frac{x^{-1/2} (1-x)^{(-1+n\epsilon)/2} }
                { [({\bf p}-{\bf k})^2 x + {\bf p}^2 (1-x) ]^{1+n\epsilon/2} }
                                                               \nonumber     \\
  & = &  \frac{1}{2 \sqrt{\pi}}  (4\pi )^{\epsilon /2}  
   \frac{\Gamma (\frac{n+1}{2}\epsilon)}{\Gamma (\frac{1+n\epsilon}{2}) }
    \frac{a_n }{|{\bf k}|^{(n+1)\epsilon}  }              \nonumber     \\
  & &  \times \int_0^1 dx \frac{1}{x^{(1+(n+1)\epsilon)/2} (1-x)^{(1+\epsilon)/2} }
\end{eqnarray} 
After performing the integral in the $x$ parameter, we find the relation 
\begin{equation}
a_{n+1} = p_{n+1} (\epsilon ) \: a_n
\end{equation}
with
\begin{equation}
p_n(\epsilon ) = 
 \frac{1}{2 \sqrt{\pi}}  (4\pi )^{\epsilon /2} 
  \frac{\Gamma (\tfrac{n\epsilon }{2}) \Gamma (\tfrac{1-n\epsilon}{2}) 
                                            \Gamma (\tfrac{1-\epsilon}{2}) }
  { \Gamma (\tfrac{1+(n-1)\epsilon}{2}) \Gamma (1-\tfrac{n + 1}{2}\epsilon) } 
\end{equation} 
In this approach, the bare vertex function can be written in compact form as 
\begin{equation}
\Gamma ({\bf 0}; {\bf k}) = \gamma_0  + \gamma_0
    \sum_{n=1}^{\infty} \lambda^n \frac{\rho^{n\epsilon} }{|{\bf k}|^{n\epsilon}  } 
                           \prod_{j=1}^{n} p_j (\epsilon )
\label{ladder}
\end{equation}
where $\rho$ is a momentum scale introduced to get the dimensionless
coupling $\lambda = \rho^{-\epsilon}\lambda_0$.

In the ladder approximation, it can be easily seen that $Z_{\psi} = 1$.
On the other hand, the renormalization factor $Z_{\psi^2}$ must have the 
general structure
\begin{equation}
Z_{\psi^2} = 1 + \sum_{n=1}^{\infty} \frac{c_n (\lambda )}{\epsilon^n}
\label{poles}
\end{equation}
The position of the different poles is determined by requiring the
finiteness of $\Gamma_{\rm ren} = Z_{\psi^2} \Gamma $ in the limit
$\epsilon \rightarrow 0$. From the expression
(\ref{ladder}), we can obtain the power series
\begin{eqnarray}
c_1 (\lambda ) & = &  -\lambda - \log(2) \: \lambda^2 
           - 2\log^2(2) \:  \lambda^3                         \nonumber   \\
  &  &   -  (\tfrac{16}{3} \log^3(2) + \tfrac{1}{8} \zeta(3)) \:  \lambda^4  
                                                               \nonumber  \\
 & & - (\tfrac{50}{3} \log^4(2) +  \log (2) \zeta(3)) \: \lambda^5  \nonumber \\
  &  &   -  (\tfrac{288}{5} \log^5(2) + 6 \log^2(2) \zeta(3) 
             + \tfrac{1}{16} \zeta(5))    \:  \lambda^6  +  \ldots  \nonumber \\
c_2 (\lambda ) & = &  \tfrac{1}{2} \: \lambda^2 + 
 \log(2) \: \lambda^3 + \tfrac{5}{2} \log^2(2) \: \lambda^4      \nonumber  \\
   &  &   + (\tfrac{22}{3} \log^3(2) + \tfrac{1}{8} \zeta(3))  \: \lambda^5 
                                                                 \nonumber  \\
 &  & + (24 \log^4(2) + \tfrac{9}{8} \log(2) \zeta(3)) \: \lambda^6  +  \ldots    
                                                                  \nonumber  \\
c_3 (\lambda ) & = & -\tfrac{1}{6} \: \lambda^3 
      - \tfrac{1}{2} \log(2) \: \lambda^4 - \tfrac{3}{2} \log^2(2) \: \lambda^5 
                                                                  \nonumber \\
 & & - (\tfrac{29}{6} \log^3(2) + \tfrac{1}{16} \zeta(3)) \: \lambda^6 + \ldots 
                                                                  \nonumber  \\
c_4 (\lambda ) & = &  \tfrac{1}{24} \: \lambda^4 
 + \tfrac{1}{6} \log(2) \: \lambda^5 + \tfrac{7}{12} \log^2(2) \: \lambda^6 
                                                     + \ldots     \nonumber  \\
c_5 (\lambda ) & = &  -\tfrac{1}{120} \: \lambda^5 
              - \tfrac{1}{24} \log(2) \: \lambda^6  + \ldots       \nonumber  \\
c_6 (\lambda ) & = &  \tfrac{1}{720} \: \lambda^6   +  \ldots
\end{eqnarray}
and so on, with the next $c_n (\lambda )$ starting each time with one more power 
of the coupling.
                                                        
Of all the poles, only the first can contribute to the anomalous dimension 
$\gamma_{\psi^2}$. This is because the theory at $d \neq 2$ has a finite limit
$\Lambda \rightarrow \infty $, and the cutoff only appears from the need to 
define the units of dimensionful quantities like $\rho$. The implicit
dependence $\lambda \sim \Lambda^\epsilon \lambda_0 $ leads to
$\Lambda  (\partial \lambda /\partial \Lambda ) = \epsilon \lambda $ and,
following Eq. (\ref{anom}),
\begin{equation}
\gamma_{\psi^2} = -\Lambda  
  \frac{\partial \lambda }{\partial \Lambda }
  \frac{\partial \log Z_{\psi^2} }{\partial \lambda }
 = - \lambda \frac{d c_1}{d \lambda }
\label{dreg}
\end{equation}
In principle, the right-hand-side of (\ref{dreg}) can contain contributions
from higher order poles in (\ref{poles}), but these will vanish provided that
$d c_{n+1}/d\lambda = c_n (d c_1/d\lambda )$, identically for every $n$ \cite{ram}.
These are key constraints in order to have a renormalizable theory, since 
they guarantee the finiteness of $\gamma_{\psi^2}$ in the limit 
$\epsilon \rightarrow 0$.
Quite remarkably, we have checked that those relations are indeed satisfied 
in our case, up to the order $\lambda^8$ for which we have computed the 
exact expression of $Z_{\psi^2} $.

\begin{figure}
\begin{center}
\mbox{\epsfxsize 7.0cm \epsfbox{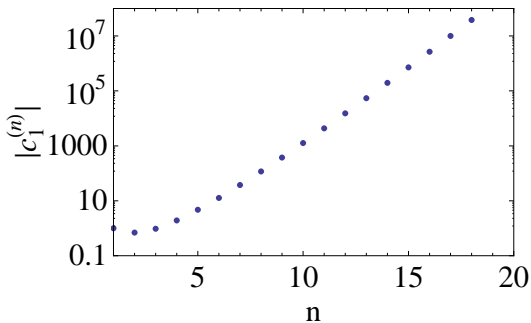}}
\end{center}
\caption{Plot of the absolute value of the coefficients $c_1^{(n)}$ in the 
expansion of $c_1 (\lambda)$ as a power series in the coupling $\lambda $.}
\label{four}
\end{figure}

The other important check we have made along the way is that
$Z_{\psi^2}$ does not contain nonlocal divergences proportional to 
$\log (|{\bf k}|/\rho )$, which appear at intermediate stages of the calculation. 
In the case of the simple pole, we have the result to order $\lambda^8$
\begin{eqnarray}
\lefteqn { c_1 (\lambda )  =   -\lambda - \log(2) \: \lambda^2 
           - 2\log^2(2) \:  \lambda^3   }                      \nonumber   \\
  &  &   -  (\tfrac{16}{3} \log^3(2) + \tfrac{1}{8} \zeta(3)) \:  \lambda^4  
 -  (\tfrac{50}{3} \log^4(2) +  \log (2) \zeta(3)) \: \lambda^5   \nonumber  \\
  &  &   -  (\tfrac{288}{5} \log^5(2) + 6 \log^2(2) \zeta(3) 
             + \tfrac{1}{16} \zeta(5))    \:  \lambda^6      
                          - (\tfrac{9604}{45} \log^6(2)           \nonumber  \\
   &  &    + \tfrac{98}{3} \log^3(2) \zeta(3)  
    + \tfrac{1}{8} \zeta^2(3)  + \tfrac{3}{4} \log(2) \zeta(5) ) \: \lambda^7   
                                                                  \nonumber  \\
  &  &  - ( \tfrac{262144}{315} \log^7(2) + \tfrac{512}{3} \log^4(2) \zeta(3) 
    + 2\log(2) \zeta^2(3)                                         \nonumber  \\ 
  &  &  +  6 \log^2(2) \zeta(5) + \tfrac{9}{256} \zeta(7) )  \:  \lambda^8  
          + O (\lambda^9 )                                                                
\label{eight}
\end{eqnarray}
The general term of this series does not have a simple expression, but one can 
still obtain numerically higher orders of the perturbative expansion 
to determine the behavior of the function $c_1 (\lambda )$. Thus, we have 
computed the coefficients $c_1^{(n)}$ of the power series in $\lambda $ up 
to order $\lambda^{18}$, what is enough to establish their exponential growth
with $n$. The results are displayed in Fig. \ref{four}, showing that
\begin{equation}
- c_1 (\lambda ) = \sum_{n=1}^{\infty} \alpha^n \lambda^n 
              + {\rm regular \;\; terms}
\end{equation}
A best fit of
the asymptotic behavior at large $n$ gives the value $\alpha \approx 4.5$.

The important point is the evidence that the anomalous dimension 
$\gamma_{\psi^2}$ obtained from (\ref{dreg}) must have a singularity at a
finite value of the effective coupling $\lambda^* = 1/\alpha$. As one approaches this 
value from below, the anomalous dimension gets arbitrarily large, meaning that 
the opening of a gap is the effect that has to prevail in the system near
$\lambda^*$, in spite of the upward renormalization of the Fermi velocity
at low energies\cite{foot}. This applies in particular to the theory with 
small number of flavors, implying that, to the right of the phase 
diagram in Fig. \ref{two}, CSB should take place above a critical coupling
$e^2/8\pi v_F = 1/\alpha \; (\approx 0.2)$.

\section{Conclusion}

In this paper we have applied renormalization group methods to analyze the 
development of a gapped phase
in graphene, taking advantage of the scaling properties of the theory 
of interacting Dirac fermions in the 2D system. 
%We have seen that, 
%in a $1/N$ approximation, the anomalous dimension of the order parameter for
%chiral symmetry breaking grows with $e^2/v_F$, though it is bounded in the 
%limit of infinite coupling constant. The field theory has then a critical 
%number of flavors $N_c$, beyond which chiral symmetry breaking does not take 
%place, no matter how large is the strength of the interaction. Furthermore, 
%we have also considered an approach suitable for small values of $N$, 
%dealing with the sum of contributions with leading logarithmic divergence. 
%We have shown that, in this ladder approximation, the anomalous dimension 
%of the order parameter grows without bound with $e^2/v_F$, so that there is 
%actually a critical coupling at which $\gamma_{\psi^2 }$ diverges.
In this regard,
an important effect that may question the breakdown of the chiral symmetry 
is the divergence of the renormalized Fermi velocity at low energies. 
In principle, the downward scaling of the effective 
coupling $e^2/v_F$ can prevent to remain above the line of the transition in
Fig. \ref{two}, even in cases where the nominal value of the coupling
places the model inside the region with $m \neq 0$. Similar objection for 
the CSB can be
applied to additional local four-fermion interactions, as their relative 
strength is always to be measured with respect to the scale of the kinetic 
energy. One may argue however that, in a statistical formulation of the 
problem, there has to be a critical temperature for the transition to 
the gapped phase. The temperature is also a relevant scale
arresting the renormalization of the Fermi velocity at low energies.
Then, it is feasible that the 
renormalized coupling $e^2/v_F$ may still keep a sufficiently large value 
to force the transition at the critical energy scale. 

We note that our results in the large-$N$ approach establish that, 
for the physical value $N = 4$, graphene would remain in the gapless phase 
even for the largest values of the effective coupling attained in vacuum 
($e^2/4\pi v_F \approx 2.2$). This is in agreement with the fact that no 
evidence of transition to an insulating state has been found in free-standing 
graphene.
The other important conclusion is that CSB must exist anyhow at sufficiently 
small values of $N$, given the evidence we have obtained of a critical coupling 
at which the anomalous dimension of the order parameter diverges. This
result could explain the observation of a transition in Monte
Carlo simulations of the long-range Coulomb interaction in the 2D 
system\cite{drut1,drut2}.

A natural prediction from our analysis is that the gapped phase should emerge 
at some point in the way from $N = 4$ to $N = 1$. The spin projection
can be frozen for instance by applying a magnetic field, and it is 
actually very appealing to think that the metal-insulator transition observed 
in that case in graphene may rest on this effect of CSB. It remains 
to be seen whether quenching also the Dirac-valley degree of freedom  
could lead to an insulating state for accessible values of $e^2/v_F$, 
in accordance with the results of this study.

{\em Acknowledgments.---}
We acknowledge financial support from MICINN (Spain) through grant
FIS2008-00124/FIS.


\begin{thebibliography}{99}
\bibitem{foot}
We note that the study in Ref. \onlinecite{fer} has also found evidence of a 
critical point in the ladder approximation, which in that case corresponds in 
general to the transition from real to complex behavior of the $\Pi$ 
susceptibility.
\end{thebibliography}
\end{document}